\documentclass[12pt,a4paper]{article}
\usepackage{amsfonts}
\usepackage{color}
\usepackage{amssymb}
\usepackage{amsmath}
\usepackage{graphicx,color}
\usepackage{amsthm}
\usepackage[latin9]{inputenc}
\usepackage{enumerate}
\usepackage{bm}
\usepackage{fancyhdr}
\usepackage{pdfsync}
\newtheorem{theor}{Theorem}

\newtheorem{lemma}{Lemma}
\newtheorem*{lemma*}{Lemma}

\numberwithin{equation}{section}
\date{}

\newcommand{\mf}{{\cal M}_F}
\newcommand{\mfz}{{\cal M}_F^{(0)}}
\newcommand{\op}{\hat{{\cal O}}}

\title{\bf {QFT over the finite line. Heat kernel coefficients, spectral zeta functions and selfadjoint extensions}}
\author{Jose. M. Mu$\tilde{\rm n}$oz-Casta$\tilde{\rm n}$eda$^1$\footnote{jose.munoz-castaneda@uni-leipzig.de}, Klaus Kirsten$^2$\footnote{klaus\_kirsten@baylor.edu} and Michael Bordag$^1$\footnote{bordag@itp.uni-leipzig.de}\\
\footnotesize{{\sl $^1$Institut f\"ur Theoretische Physik, Universit\"at Leipzig, Leipzig, 04103, Germany.}}\\
\footnotesize{{\sl $^2$GCAP-CASPER Department of Mathematics, Baylor University, Waco, TX 76798, USA.}}}

\begin{document}

\maketitle

\begin{abstract}
Following the seminal works of Asorey-Ibort-Marmo and Mu\~{n}oz-Casta\~{n}eda-Asorey about selfadjoint extensions and quantum fields in bounded domains, we compute all the heat kernel coefficients for any
strongly consistent selfadjoint extension of the Laplace operator over the finite line $[0,L]$. The derivative of the corresponding spectral zeta function at $s=0$ (partition function of the corresponding quantum field theory) is obtained. In order
to compute the correct expression for the $a_{1/2}$ heat kernel coefficient, it is necessary
to know in detail which non-negative selfadjoint extensions have zero modes and how many of them they have. The answer to this question
leads us to analyse zeta function properties for the Von Neumann-Krein extension, the only extension with two zero modes.

\end{abstract}

\section{Introduction. Basic formulas and results}

The physical system on which we will focus is a free massless scalar quantum field theory defined over the finite interval $[0,L]$. The quantum Hamiltonian that describes the one particle states of this quantum field theory is given by the Laplace operator over the finite line $[0,L]$. It is a very well known fact that the Laplace operator over the finite line $[0,L]$ is not an essentially selfadjoint operator but instead admits an infinite set of selfadjoint extensions. We will denote by ${\cal M}$ the set of all the selfadjoint extensions of $\Delta$ over the finite line $[0,L]$. Physically speaking this means that there is an infinite set of possible quantum field theories that describe the behavior of a free quantum massless scalar field confined to propagate in the interval $[0,L]$. In order to respect the unitarity principle of quantum field theory we must only take into account those selfadjoint extensions of the Laplace operator that give rise to non-negative selfadjoint operators (see \cite{asor13-874-852}). As described in \cite{asor13-874-852} among the set of non-negative selfadjoint extensions we can distinguish between two different types:
\begin{enumerate}
\item Non-negative selfadjoint extensions of $\Delta$ over $[0,L]$ that are non-negative only for certain values of the finite length $L$ of the interval. Typically these selfadjoint extensions are non-negative for $L\geq L_0$ for a given $L_0$ that depends on the selfadjoint extension. When $L<L_0$ these selfadjoint extensions have negative eigenvalues and thus give rise to non unitary quantum field theories. We will call these selfadjoint extensions {\it weakly consistent selfadjoint extensions}.

\item Non-negative selfadjoint extensions of $\Delta$ over $[0,L]$ that are non-negative for any value of the finite length $L$ of the finite line. These selfadjoint extensions have only zero and positive eigenvalues for any value of $L\in(0,\infty)$. We will call these selfadjoint extensions {\it strongly consistent selfadjoint extensions} and following \cite{asor13-874-852} we denote by ${\cal M}_F$ the set of strongly consistent selfadjoint extension.
\end{enumerate}

In this paper we will focus only on those free massless scalar quantum field theories defined by selfadjoint extensions of $\Delta$ over $[0,L]$ that are non-negative for all $L\in(0,\infty)$. Hence we will only be interested in the selfadjoint extensions contained in ${\cal M}_F$. This restriction is very natural from a quantum field theoretical point of view because the space ${\cal M}_F$ is stable under the renormalization group transformations. On the other hand the whole set of non-negative selfadjoint extensions for fixed length $L$ is not stable under renormalization group transformations since there are non-negative selfadjoint extensions that will loose the non-negativity condition under a renormalization group transformation giving rise to non-unitary quantum field theories (see Ref. \cite{aso-jpa07}). Typically one distinguishes separated and coupled boundary conditions \cite{zett05b}, but in the formulation
of \cite{asor13-874-852} this will not be necessary (an extension of the AIM formalism was first addressed in the first chapter of Ref. \cite{muca-phd09} and later on reformulated in a more rigorous approach in Ref. \cite{perez-pardo}).

In order to be able to characterize the selfadjoint extensions of ${\cal M}_F$ we will use the Asorey-Ibort-Marmo (AIM) formalism (see \cite{asor05-20-1001}) to characterize the selfadjoint extensions of $\Delta$ over the finite line $[0,L]$. From the first AIM theorem (see \cite{asor05-20-1001,asor13-874-852}) the set of selfadjoint extensions of $\Delta$ over $[0,L]$ is in one-to-one correspondence with the group ${\rm U}(2)$. Given any $U\in{\rm U}(2)$ we will denote the corresponding selfadjoint extension by $\Delta_U$. Each selfadjoint extension $\Delta_U$ is defined by its domain of functions ${\cal D}_U\subset H^2([0,L],\mathbb{C})$ being $H^2([0,L],\mathbb{C})$ the Sobolev space of functions over the finite interval that are $L^2$ together with their derivatives up to second order. The domain ${\cal D}_U\subset H^2([0,L],\mathbb{C})$ that defines the selfadjoint extension $\Delta_U$ is given in terms of the matrix $U\in{\rm U}(2)$  (see \cite{asor05-20-1001,asor13-874-852}) by
\begin{equation}
   {\cal D}_U=\left\{ \psi\in H^2([0,L],\mathbb{C})/ \,\, \varphi-i\dot\varphi=U(\varphi+i\dot\varphi)\right\},\label{dom}
\end{equation}
where $\varphi$ and $\dot\varphi$ are the boundary data\footnote{It is worth noting that for spacetime dimension higher than $1+1$ the maximal domain of the symmetric operator $\Delta$ defined over the compact manifold $M$ with smooth boundary $\partial M$ is the Sobolev space $H^2(M,\mathbb{C})$ and the domain of its adjoint $\Delta^\dagger$ is $H^2(M,\mathbb{C})\oplus\ker(\Delta^\dagger)$. This is of crucial importance from a physical point of view because boundary values of the fields that have singularities represent important physical situations as for example point charge distributions over the boundary. Nevertheless in the case of spacetime dimension $1+1$ and $M=[0,L]$ things become simpler because the space of boundary data is the linear vector space $\mathbb{C}^{2}$ (see Refs. \cite{muca-phd09,perez-pardo}). } for $\psi\in H^2([0,L],\mathbb{C})$:
\begin{equation}
  \varphi\equiv\left(\begin{tabular}{c} $\psi(0)$ \\ $\psi(L)$ \end{tabular}\right),\quad \dot\varphi\equiv\left(\begin{tabular}{c} $-\psi'(0)$ \\ $\psi'(L)$ \end{tabular}\right).
\end{equation}
Following the notation in \cite{asor13-874-852} for any  $\psi\in H^2([0,L],\mathbb{C})$ we introduce the 2 dimensional column vectors $\varphi_\pm(\psi)$:
\begin{equation}
   \varphi_\pm(\psi)\equiv\left(\begin{tabular}{c} $\psi(0)\mp i\psi'(0)$ \\ $\psi(L)\pm i\psi'(L)$ \end{tabular}\right).\label{phipm}
\end{equation}
We can write the boundary condition given in eq. (\ref{dom}) in terms of $\varphi_\pm(\psi)$  as
\begin{equation}
   \varphi_-(\psi)=U\cdot\varphi_+(\psi) .\label{bcphipm}
\end{equation}
Following the conventions and notation used in \cite{asor13-874-852} we parameterize the elements $U\in{\rm U}(2)$ by using 5 parameters:
\begin{equation}
U(\alpha,\beta,{\bf n})=e^{i\alpha}\left[ \cos(\beta)\mathbb{I}+i\sin(\beta) ({\bf n}\cdot \boldsymbol{\sigma})\right], \label{uparam}
\end{equation}
where $\mathbb{I}$ is the 2$\times$2 identity matrix, $\boldsymbol{\sigma}=(\sigma_1,\sigma_2,\sigma_3)$ are the Pauli matrices, ${\bf  n}$ is a 3 dimensional unit vector ($n_1^2+n_2^2+n_3^2=1$) and the angles $\alpha$ and $\beta$ are such that
\begin{equation}
\alpha\in [-\pi,\pi];\quad\beta\in[-\pi/2,\pi/2].
\end{equation}
Using this parametrization we can characterize the non-zero part of the spectrum for any $\Delta_U\in{\cal M}$ including multiplicities of eigenvalues
(see the consistency lemma in \cite{asor13-874-852}). The  secular equation
obtained in \cite{asor13-874-852} for any $\Delta_U\in{\cal M}$ is given by
\begin{eqnarray}
h_U(k)&=&\nonumber 2i e^{i\alpha}\left[ \sin(kL)\left((k^2-1)\cos(\beta)+(k^2+1)\cos(\alpha)\right)\right.\\
&-&\left.2k\sin(\alpha)\cos(kL)-2k n_1\sin(\beta)\right].\label{hspec}
\end{eqnarray}
The non-zero part $\tilde\sigma(\Delta_U)$ of the spectrum of $\Delta_U\in{\cal M}$ is given by
\begin{equation}
  \tilde\sigma(\Delta_U)=\{k^2\in\mathbb{R}-\{0\}/\,\,h_U(k)=0\}=\{k^2\in\mathbb{R}-\{0\}/\,\,k\in{\rm Z}(h_U)-\{0\}\},
\end{equation}
where ${\rm Z}(h_U)$ denotes the set of zeroes of the function $h_U(k)$. For any non-zero root of $h_U(k)$ the multiplicity $d_U(k^2)$ of the corresponding eigenvalue is
\begin{equation}
\forall k\in{\rm Z}(h_U)-\{0\}:\quad d_U(k^2)=\left.{\rm Res}\left(\frac{d}{dz}\log(h_U(z))\right)\right|_{z=k}.
\end{equation}
Let us mention, that the bound states of a given selfadjoint extension $\Delta_U\in{\cal M}$ are given by zeroes of $h_U(z)$ of the form $k=i\kappa$ with $\kappa>0$, i.e. $k^2<0$. Furthermore, note that from eq. (\ref{hspec}) it is easy to see that $\lim_{k\to 0}h_U(k)=0$. This fact does not ensure that the corresponding selfadjoint extension $\Delta_U$ admits zero modes. The question about which selfadjoint extensions of ${\cal M}_F$ admit zero modes will be solved in the next section.
\par
\begin{figure}[h]
\center{\includegraphics[width=6cm]{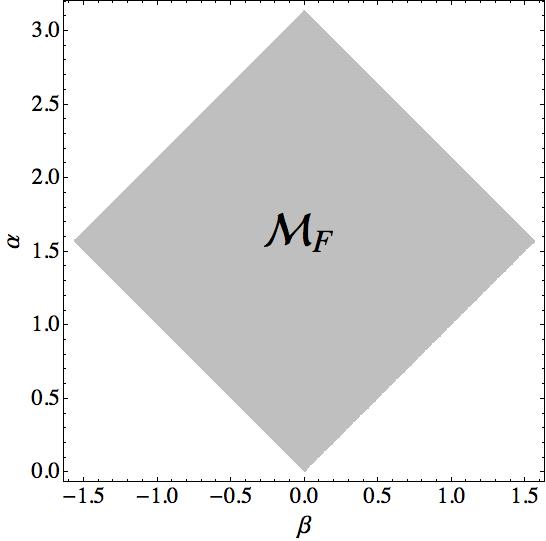}}
\caption{\footnotesize{This graphic shows the set ${\cal M}_F$ in the $\alpha\beta$-plane. In the top corner of the rhombus are placed the Dirchlet boundary conditions, the bottom corner corresponds to Neumann boundary conditions meanwhile the left and right corners correspond to periodic (left for $n_1=1$ and right for $n_1=-1$) and anti-periodic (left for $n_1=-1$ and right for $n_1=1$).}}
\label{mfpic}
\end{figure}

Once all the selfadjoint extensions of ${\cal M}$ have been explicitly characterized using the AIM formalism (see \cite{asor05-20-1001} for details), following \cite{asor13-874-852} we can characterize all the selfadjoint extensions that belong to ${\cal M}_F$ and hence that give rise to strongly consistent quantum field theories\footnote{Given that the AIM formalism (first AIM theorem in \cite{asor05-20-1001,asor13-874-852})
ensures the one-to-one correspondence between selfadjoint extensions of $\Delta$ over $[0,L]$ and unitary matrices of ${\rm U}(2)$ from now on we will not make a distinction between selfadjoint extensions $\Delta_U\in{\cal M}$ and unitary matrices $U\in{\rm U}(2)$ (see the appendix)}. One of the main results in \cite{asor13-874-852} is the characterization of the set ${\cal M}_F\subset{\cal M}$ of non-negative selfadjoint extensions $\forall L\in(0,\infty)$ (``strong consistency lemma''):
\begin{equation}
{\cal M}_F=\{U(\alpha,\beta,{\bf n})\in{\rm U}(2)={\cal M}/\quad 0\leq\alpha\pm\beta\leq\pi\}.\label{mf}
\end{equation}
In Figure \ref{mfpic} we can see a representation of the set ${\cal M}_F$ in the $\alpha\beta$-plane.
Whereas extensive results on the spectral zeta functions and the heat kernel are available for the standard boundary conditions like Dirichlet, Neumann, Robin or periodic \cite{eliz94b,gilk95b,gilk04b,kirs02b,vass03-388-279},
general boundary conditions as described in (\ref{bcphipm}) have not been analyzed in comparable detail. This is the topic of the current paper.
Generic interest in the analysis of spectral functions stems from their relevance in global analysis \cite{gilk95b,ray71-7-145}
and quantum field theory topics such as the Casimir effect
\cite{blau88-310-163,bord09b,bord01-353-1,byts96-266-1,dowk76-13-3224,dowk78-11-895,hawk77-55-133,milt01b}.

The paper is organized as follows.
In Section 2 we will answer the question which selfadjoint extensions within the strongly consistent extensions allow for zero modes. This is necessary as the details of the zeta function
analysis depend on this input. Based upon the function $h_U (k)$, eq. (\ref{hspec}), a contour integral representation of the zeta function for any strongly consistent
selfadjoint extension will be derived. As usual, residues and certain values of the zeta function determine the associated heat kernel coefficients. The cases with and without zero modes are
treated in different subsections of Section 3. Results for standard boundary conditions are verified as a check. In Section 4 we use the integral representation of the zeta function to
compute its derivative at $s=0$, once again for all possible cases. Checks for known results
are provided. In the conclusions we summarize the most important aspects of our work together with possible future directions of research.

\section{Zero modes of $\Delta_U\in{\cal M}_F$}

The purpose of this section is to study the zero mode structure of selfadjoint extensions contained in ${\cal M}_F$. In particular we will focus our attention on two main questions:
\begin{itemize}
  \item Characterize the subset ${\cal M}^{(0)}_F\subset{\cal M}_F$ of selfadjoint extensions that have zero modes,
  \begin{equation}
  {\cal M}^{(0)}_F\equiv\left\{ \Delta_U\in{\cal M}_F/\quad 0\in\sigma(\Delta_U)\right\}.
  \end{equation}

  \item Study the zero mode structure and compute $\dim\left(\ker \Delta_U\right)$ of any $\Delta_U\in{\cal M}^{(0)}_F$.
\end{itemize}
The motivation to study these two questions about the zero modes of the selfadjoint extensions contained in ${\cal M}_F$ is to obtain a correct result of the $a_{1/2}$ heat kernel coefficient, for which we must know explicitly $\dim\left(\ker \Delta_U\right)$.
There are no contributions of zero modes to residues of the zeta function.
\par
The differential equation for the zero modes is
\begin{equation}
\frac{d^2}{dx^2}\psi_0(x)=0,\label{eqzm}
\end{equation}
and its general solution is given by
\begin{equation}
\psi_0(x)=a+b x,\label{genzm}
\end{equation}
where $a$ and $b$ are complex constant numbers. Notice that:
\begin{itemize}
  \item When $\Delta$ is defined over the whole real line, the only solution to eq. (\ref{eqzm}) given by (\ref{genzm}) with finite ${\cal L}^2$ norm is given by $a=b=0$. Hence when $\Delta$ is defined over the real line there are no zero modes.

  \item On the other hand, when $\Delta$ is defined over the finite line $[0,L]$, due to the finite length of the interval the general solution (\ref{genzm}) has always finite ${\cal L}^2$ norm. Hence when $\Delta$ is defined over the finite interval there exists the possibility of having constant and linear zero modes.
\end{itemize}
Given a selfadjoint extension $\Delta_U\in{\cal M}_F$, in order to decide if it admits zero modes of the general form (\ref{genzm}) we must impose over (\ref{genzm}) the corresponding boundary condition given by (\ref{dom}). From eq. (\ref{phipm}) we obtain for $\psi_0(x)$
\begin{equation}
   \varphi_\pm(\psi_0)\equiv\left(\begin{tabular}{c} $a\mp ib$ \\ $a+b(L\pm i)$ \end{tabular}\right)= \left(\begin{tabular}{cc} 1 & $\mp i$ \\ 1 & $L\pm i$ \end{tabular}\right)\cdot \left(\begin{tabular}{c} $a$ \\ $b$ \end{tabular}\right).\label{phizm}
\end{equation}
Using (\ref{phizm}) in the boundary condition (\ref{bcphipm}) we obtain the linear system
\begin{equation}
  \left[\left(\begin{tabular}{cc} 1 & $i$ \\ 1 & $L- i$ \end{tabular}\right)-U\cdot \left(\begin{tabular}{cc} 1 & $- i$ \\ 1 & $L+ i$ \end{tabular}\right)\right]\cdot \left(\begin{tabular}{c} $a$ \\ $b$ \end{tabular}\right)=0.\label{bczm}
\end{equation}
This linear system is nothing else than the boundary condition for the zero modes. Given its importance in this section, let us call the matrix of the linear system $D_U$:
\begin{equation}
  D_U=\left(\begin{tabular}{cc} 1 & $i$ \\ 1 & $L- i$ \end{tabular}\right)-U\cdot \left(\begin{tabular}{cc} 1 & $- i$ \\ 1 & $L+ i$ \end{tabular}\right) .\label{du}
\end{equation}
Next we investigate the solutions of the linear system (\ref{bczm}).

\subsection{The first question: characterization of ${\cal M}_F^{(0)}$}
From basic algebra we know that $\Delta_U\in{\cal M}_F$ will admit zero modes if and only if the linear system (\ref{bczm}) has non-trivial solutions, i.e.
\begin{equation}
  \ker(\Delta_U)\neq 0\,\,\Leftrightarrow\,\,\ker (D_U)\neq 0\,\,\Leftrightarrow\,\,\det(D_U)=0.
\end{equation}
Hence the characterization of ${\cal M}_F^{(0)}$ is given by
\begin{equation}
  {\cal M}_F^{(0)}=\left\{ U\in{\cal M}_F/\quad \det(D_U)=0\right\}.
\end{equation}
To explicitly compute all the selfadjoint extensions contained in ${\cal M}_F^{(0)}$ we need to solve the secular equation of the linear system (\ref{bczm})
\begin{equation}
  \det(D_U)=0 . \label{eqdu}
\end{equation}
Introducing the parametrization (\ref{uparam}) in (\ref{du}) and simplifying we obtain
\begin{equation}
  \det(D_U)=2 e^{i\alpha}\left[ L\left( \cos(\alpha)-\cos(\beta)\right)-2\left( \sin(\alpha)+n_1\sin(\beta)\right)\right].\label{detdu}
\end{equation}
Therefore neglecting the global factor $2 e^{i\alpha}$ that is never zero the equation to solve is
\begin{equation}
   L\left( \cos(\alpha)-\cos(\beta)\right)-2\left( \sin(\alpha)+n_1\sin(\beta)\right)=0\label{seceq}
\end{equation}
with the restrictions ensuring that the corresponding solution gives a matrix $U$ that is in ${\cal M}_F$:
\begin{equation}
  n_1\in[-1,1];\,\,\alpha\in[0,\pi];\,\,\beta\in[-\pi/2,\pi/2];\,\, 0\leq\alpha\pm\beta\leq\pi  \,.\label{mfcond}
\end{equation}
The simplest way to solve (\ref{seceq}) is by imposing $$\cos \alpha - \cos \beta =0 \Longrightarrow \alpha = \pm \beta,$$
which makes $$\sin \alpha + n_1 \sin \beta = \sin \alpha \pm n_1 \sin \alpha =0 \Longrightarrow n_1 = \mp 1$$ necessary. In fact,
these are all possible solutions. Because if $\cos \alpha - \cos \beta \neq 0$, we have
$$ L = \frac{ 2\sin \alpha + n_1 \sin \beta }{ \cos \alpha - \cos \beta } $$
with $L>0$. However, with the parameters confined by the conditions in (\ref{mfcond}) one can show that the right hand side is always negative.
As a consequence we have shown that
all the solutions to (\ref{seceq}) that satisfy conditions (\ref{mfcond}) are given by
\begin{equation}
   {\cal M}_F^{(0)}=\left\{U\in{\cal M}_F/\quad n_1=\pm 1;\,\,\alpha\in[0,\pi/2];\,\,\beta=-n_1\alpha\right\}  .  \label{premf0}
\end{equation}

\begin{figure}[h]
\center{\includegraphics[width=6cm]{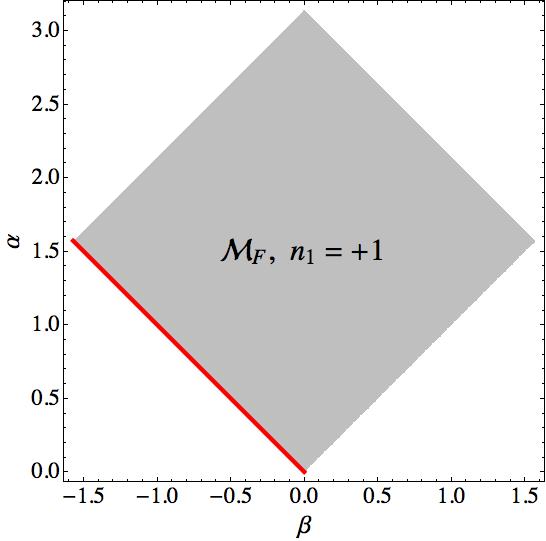}\qquad\includegraphics[width=6cm]{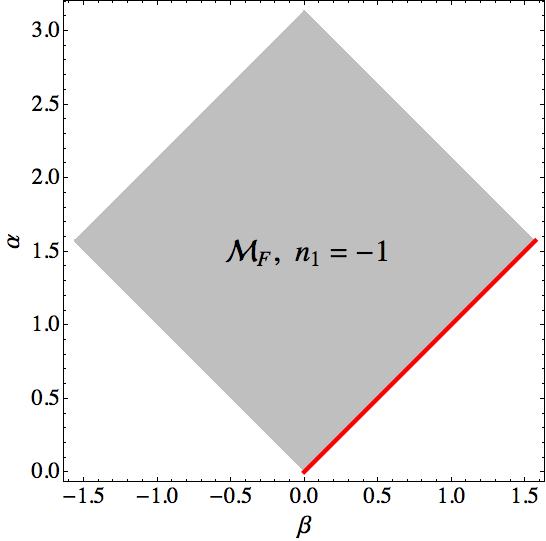}}
\caption{\footnotesize{Representation of ${\cal M}_F^{(0)}$ (red lines) over the $\alpha\beta$-plane}}
\label{mfzeropic}
\end{figure}

\par
In terms of the parametrization given in (\ref{uparam}) the unitary matrices contained in ${\cal M}_F^{(0)}$ are given by
\begin{equation}
   U\in{\cal M}_F^{(0)}\Rightarrow U=e^{i\alpha}\left[ \cos(\alpha)\mathbb{I}-i n_1\sin(\alpha)\sigma_1 \right];\,\,\alpha\in[0,\pi/2];\,\,n_1\in \{-1,1\}. \label{umfz}
\end{equation}
Using the expression above for the matrices contained in ${\cal M}_F^{(0)}$ and the definition (\ref{du}), we find
\begin{equation}
   D_U=\left(\begin{tabular}{cc} 0 & $i e^{i\alpha}\left(2\cos(\alpha)+L\sin(\alpha)\right)$ \\ 0 & $-i e^{i\alpha}\left(2\cos(\alpha)+L\sin(\alpha)\right)$ \end{tabular}\right)\quad\forall \,\,U\in{\cal M}_F^{(0)}.\label{dumfz}
\end{equation}
As can be seen from this expression above, when $U\in{\cal M}_F^{(0)}$ the matrix $D_U$ has indeed zero determinant. In Figure \ref{mfzeropic} it is shown the space ${\cal M}_F^{(0)}$ in the $\alpha\beta$-plane.

\subsection{The second question: $\dim\left(\ker(\Delta_U)\right)$ for $\Delta_U\in{\cal M}_F^{(0)}$}

Taking into account eq. (\ref{bczm}) and the meaning of the constants $a$ and $b$, see eq. (\ref{genzm}),
the second question will be answered by studying the explicit solutions to (\ref{bczm}) when $D_U$ is given by expression (\ref{dumfz}), i.e. $U\in{\cal M}_F^{(0)}$.
We will answer this second question in two lemmas with their corresponding demonstrations.

\begin{lemma}\label{le1}
Any selfadjoint extension $\Delta_U\in{\cal M}_F^{(0)}$ admits a constant zero mode.
\end{lemma}
\paragraph{Proof.} To proof the lemma we only need to demonstrate that the column vector
\begin{equation}
v_c^{(0)}=\left(\begin{tabular}{c} a  \\ 0 \end{tabular}\right),\quad a\neq 0,\label{vc}
\end{equation}
belongs to $\ker(D_U)$ for any $U\in{\cal M}_F^{(0)}$ (notice that according to (\ref{genzm}) when $b=0$ and $a\neq0$ the expression gives rise to the constant function over the interval $[0,L]$). For any $U\in{\cal M}_F^{(0)}$ the associated matrix $D_U$ is given by (\ref{dumfz}). Since the first column in (\ref{dumfz}) is identically zero by direct trivial calculation
\begin{equation}
   D_U\cdot v_c^{(0)}=\left(\begin{tabular}{c} 0  \\ 0 \end{tabular}\right)\quad \forall\,\,U\in{\cal M}_F^{(0)}.
\end{equation}
Therefore $v_c^{(0)}$ is a solution to the linear system (\ref{bczm}). Hence taking (\ref{genzm}) into account for any $\Delta_U\in{\cal M}_F^{(0)}$ there exists a constant zero mode. $\blacksquare$

This lemma ensures that any selfadjoint extension $\Delta_U\in{\cal M}_F^{(0)}$ has at least a constant zero mode, i.e.
\begin{equation}
   \forall\,\,\Delta_U\in{\cal M}_F^{(0)},\quad \,\,\,\dim\left(\ker(\Delta_U)\right)=\dim\left(\ker(D_U)\right)\geq 1.\label{dimker1}
\end{equation}
Since any  $\Delta_U\in{\cal M}_F^{(0)}$ has a constant zero mode the only possibility to be explored now is the possibility of having selfadjoint extensions  $\Delta_U\in{\cal M}_F^{(0)}$ that also admit a linear zero mode.
The condition for a selfadjoint extension  $\Delta_U\in{\cal M}_F^{(0)}$ to admit a linear zero mode is given by
\begin{equation}
   \dim\left(\ker(\Delta_U)\right)=\dim\left(\ker(D_U)\right)=2.\label{dimker2}
\end{equation}
Since $D_U$ is a $2\times 2$ complex matrix
\begin{equation}
   \dim\left(\ker(D_U)\right)=2\,\,\Longleftrightarrow\,\, D_U=0.\label{duzero}
\end{equation}
This condition ensures the existence of a linear zero mode for any selfadjoint extension $\Delta_U\in{\cal M}_F^{(0)}$ by the following argumentation:
\begin{enumerate}[i.]
  \item For any  $\Delta_U\in{\cal M}_F^{(0)}$ there is a constant zero mode $\Rightarrow\,\,v_c^{(0)}$ given by (\ref{vc}) belongs to $\ker(D_U)$ for any $\Delta_U\in{\cal M}_F^{(0)}$.
  \item $\Delta_U\in{\cal M}_F^{(0)}$ will admit a linear zero mode if and only if the matrix $D_U$ is such that there exists in addition to $v_c^{(0)}$ a solution to the linear system (\ref{bczm}) with $b\neq 0$ (see eq. (\ref{genzm})).
  \item Hence $\Delta_U\in{\cal M}_F^{(0)}$ will admit a linear zero mode if and only if
  \begin{equation}
     \dim\left(\ker(D_U)\right)=2\,\,\Longleftrightarrow\,\, D_U=0,
  \end{equation}
  because any solution to (\ref{bczm}) with $b\neq 0$ will be linearly independent of the vector $v_c^{(0)}\in\ker(D_U)\,\,\forall\,\, \Delta_U\in{\cal M}_F^{(0)}$.
\end{enumerate}

\begin{lemma}\label{le2}
There are no selfadjoint extensions $\Delta_U\in{\cal M}_F^{(0)}$ that admit a linear zero mode.
\end{lemma}
\paragraph{Proof.} Given any $\Delta_U\in{\cal M}_F^{(0)}$ the necessary and sufficient condition to admit a linear zero mode is (\ref{duzero}). Since for $\Delta_U\in{\cal M}_F^{(0)}$ the associated $D_U$ matrix is given by (\ref{dumfz}) the condition $D_U=0$ is given by the equation
\begin{equation}
   2\cos(\alpha)+L\sin(\alpha)=0\,\,\Rightarrow\,\,\tan(\alpha)=-2/L.\label{dueqz}
\end{equation}
Because $L$ is the length of the interval $-2/L\leq 0$. Therefore there is no $\alpha\in[0,\pi/2]$ satisfying (\ref{dueqz})\footnote{Since $\Delta_U\in{\cal M}_F^{(0)}$ the angle $\alpha$ is restricted to lie in the interval $[0,\pi/2]$.}. Therefore no $\Delta_U\in{\cal M}_F^{(0)}$ can satisfy the condition $D_U=0$, i.e. no $\Delta_U\in{\cal M}_F^{(0)}$ admits a linear zero mode. $\blacksquare$

To conclude this section we compile all the results in the following theorem.
\begin{theor}
  The space ${\cal M}_F^{(0)}\subset{\cal M}_F$ of non-negative selfadjoint extensions of the Laplace operator $\Delta$ over $[0, L]$ that admit zero modes is given by
  \begin{equation}
   {\cal M}_F^{(0)}=\left\{U\in{\cal M}_F/\quad n_1=\pm 1;\,\,\alpha\in[0,\pi/2];\,\,\beta=-n_1\alpha\right\}.\label{mf0}
\end{equation}
In addition $\dim\left(\ker(\Delta_U)\right)=1$ for any selfadjoint extension $\Delta_U\in{\cal M}_F^{(0)}$ and the unique zero mode is the constant function over the interval $[0,L]$.
\end{theor}

\subsection{A remark about the Von Neumann-Krein extension}
To complete the study of the zero-modes we will determine the minimal non-negative selfadjoint extension: the so-called Von Neumann-Krein extension. To introduce the general definition of the Von Neumann-Krein extension we need a quick overview of some general results (see Refs. \cite{alsi-jot80,teschl-am10,teschl-12}). Let $T_1$ and $T_2$ be two non-negative selfadjoint operators with dense domains in a Hilbert space ${\cal H}$. We say that
\begin{equation}
T_1\leq T_2
\end{equation}
if and only if
\begin{enumerate}[i]
\item ${\cal D}(T_2^{1/2})\subseteq {\cal D}(T_1^{1/2}),$
\item $\langle T_1^{1/2}\psi\vert T_1^{1/2}\psi \rangle_{L^2}\leq \langle T_2^{1/2}\psi\vert T_2^{1/2}\psi \rangle_{L^2}$ for all $\psi\in {\cal D}(T_2^{1/2}).$
\end{enumerate}

Let now $T$ be a non-negative symmetric operator over a Hilbert space ${\cal H}$. Then there exist two unique non-negative selfadjoint extensions $T_{min}$ and $T_{max}$ such that $T_{min}\leq T_{max}$ and every non-negative selfadjoint extension $S$ of $T$ satisfies
\begin{equation}
T_{min}\leq S\leq T_{max}.
\end{equation}
The minimal non-negative selfadjoint extension $T_{min}$ is the so-called Von Neumann-Krein (VNK) extension. From now on we will denote the Von Neumann-Krein extension with the sub-index $VNK$.

Following subsection 11.1 in Ref. \cite{teschl-12} the VNK extension of the operator $T=-\Delta$ over the finite interval $[0,L]$ is characterized as the unique selfadjoint extension with a maximal number of zero modes. From
eq. (\ref{bczm}) the maximum number of zero-modes for the Laplace operator over the finite line is two: a constant zero-mode, and a linear zero mode. Therefore the condition that characterizes uniquely the VNK extension is
\begin{equation}
D_U=0 \Rightarrow U_{VNK}=\left(\begin{tabular}{cc} 1 & $i$ \\ 1 & $L- i$ \end{tabular}\right)\cdot \left(\begin{tabular}{cc} 1 & $- i$ \\ 1 & $L+ i$ \end{tabular}\right)^{-1},
\end{equation}
\begin{equation}
 U_{VNK}=\frac{1}{L+2i}\left(\begin{tabular}{cc} $L$ & $2i$ \\ $2 i$ & $L$ \end{tabular}\right).\label{uvnk}
\end{equation}
It is straightforward to check that $U_{VNK}$ is a unitary matrix and therefore defines a selfadjoint extension of the Laplacian over the finite line. In order to demonstrate
whether or not the VNK extension belongs to ${\cal M}_F$ we must compute the parameters $\{\alpha_{VNK},\beta_{VNK},{\bf n}_{VNK}\}$ that characterize the VNK extension in the parametrization given by (\ref{uparam}). Since (\ref{uvnk}) is a symmetric matrix and both diagonal elements are equal we must require $n_2=n_3=0$. Therefore we can assume without loss of generality that ${\bf n}_{VNK}=(1,0,0)$. Knowing that
\begin{equation}
\left.{\bf U}\right\vert_{n_1=1}=e^{i\alpha}\left(\begin{tabular}{cc} $\cos(\beta)$ & $i\sin(\beta)$ \\ $ \sin(\beta)i$ & $\cos(\beta)$ \end{tabular}\right)
\end{equation}
and comparing with (\ref{uvnk}) we obtain the following two equations:
\begin{eqnarray}
e^{i\alpha}\cos(\beta)&=&\frac{L}{L+2i},\label{A}\\
e^{i\alpha}\sin(\beta)&=&\frac{2}{L+2i}.\label{B}
\end{eqnarray}
Dividing eq.~(\ref{B}) by eq.~(\ref{A}) it follows that $\tan(\beta_{VNK})=2/L$. Since the principal value of $\arctan(x)$ is in the interval $[-\pi/2,\pi/2]$ we conclude that $\beta_{VNK}=\arctan\left(2/L\right)$.
In addition it is easy to see that $\sin(\beta_{VNK})=2/\sqrt{L^2+4}$ and $\cos(\beta_{VNK})=L/\sqrt{L^2+4}$. To determine $\alpha_{VNK}$ we sum (\ref{A})$+i$(\ref{B}) to obtain the equation $e^{i(\alpha_{VNK}+\beta_{VNK})}=1\Rightarrow\alpha_{VNK}=-\beta_{VNK}=-\arctan\left(2/L\right)$. Hence the VNK extension is characterized by:
\begin{eqnarray}
&& {\bf n}_{VNK}=(1,0,0),\quad \alpha_{VNK}(L)=-\beta_{VNK}(L),\\
&&\beta_{VNK}(L)=\arctan\left(2/L\right).
\end{eqnarray}
Taking into account that $\arctan\left(2/L\right)$ is a non-negative and monotonically decreasing function in the interval $L\in[0,\infty)$ that goes from the value $\pi/2$ for $L\rightarrow0$ to the value $0$ when $L\rightarrow\infty$ we conclude that $\beta_{VNK}\in[0,\pi/2]$ and $\alpha_{VNK}\in[-\pi/2,0]$ for any $L\in(0,\infty)$. Therefore $U_{VNK}\notin{\cal M}_F$ for any value of $L$. Nevertheless, since the VNK extension is non-negative we will be able to compute the heat kernel coefficients and the derivative at zero of the spectral zeta function with the methods we develop in the following sections.

\section{The heat kernel expansion of $\Delta_U\in{\cal M}_F$}

Using standard methods described for example in Ref. \cite{kirs02b} we will next compute all the coefficients of the asymptotic expansion of the heat kernel corresponding to any selfadjoint extension $\Delta_U\in{\cal M}_F$. Before going over the explicit calculation let us introduce the general results contained in \cite{kirs02b} that will be necessary in our calculation.

Let $\op$ be an elliptic non-negative selfadjoint second order differential operator (in one dimension) over a Hilbert space ${\cal H}$. Let $f_{\op} (z)$ be a holomorphic function over the complex plane such that for $k\in\mathbb{R}$
\begin{equation}
\lim_{k\rightarrow 0}f_{\op}(k)\neq 0,\,\infty,\label{specfzero}
\end{equation}
and such that the non-zero part of the spectrum of $\op$ is given by\footnote{We will denote by $\sigma(\op)$ the spectrum of the operator $\op$ and $\tilde{\sigma}(\op)$ the non zero part of $\sigma(\op)$. Given a function $f(z)$ over the complex plane we will denote by $Z(f)$ the set of its zeroes over the complex plane.}
\begin{equation}
  \tilde{\sigma}(\op)=Z(f_{\op}),
\end{equation}
where the multiplicities of eigenvalues are reflected in the order of the zeroes.
When $f_{\op}$ satisfies the conditions stated above, the spectral zeta function of the operator $\op$ can be written as:
\begin{equation}
  \zeta_{\op}(s)=\frac{\sin(\pi s)}{\pi}\int_0^\infty dk\cdot k^{-2 s}\partial_k\log\left(f_{\op}(i k)\right).\label{zetagen}
\end{equation}
This approach has been used for many examples; see, e.g., \cite{kirs02b,kirs03-308-502}.
The integral in (\ref{zetagen}) in the current context will be convergent in the region $1/2 < \Re s <1$.
However, expression (\ref{zetagen}) admits an analytical continuation to the whole complex plane with, in general, poles at
\begin{equation}
  s=\frac{1}{2}-n;\quad n=0,\, 1,\, 2,\, 3...\label{genzpoles}
\end{equation}
The heat kernel coefficients can be computed in terms of the residues at the poles and the values at non-positive integers of $\zeta_{\op}(s)$ \cite{seel68-10-288}:
\begin{equation}
  a_{1/2-z}(\op)=\Gamma(z){\rm Res}\left(\zeta_{\op},s=z\right),\label{genhc1}
\end{equation}
\begin{equation}
  a_{1/2+q}(\op)=(-1)^q\frac{\zeta_{\op}(-q)}{\Gamma(q+1)}+\delta_{q,0}N_Z(\op).\label{genhc2}
\end{equation}
In eq. (\ref{genhc2}), $N_Z(\op)$ denotes the number of zero modes of the operator $\op$.

Hence, according to formulas (\ref{genhc1}) and (\ref{genhc2}), in order to know all the heat kernel coefficients we only need to know the residues at the poles and the values at the
non-positive integers of the spectral zeta function $\zeta_{\op}(s)$. To use formula (\ref{zetagen}) we will need to use the secular
equation given in formula (\ref{hspec}). Note, however, that $k$ needs to be replaced by $ik$ when used in (\ref{zetagen}).

Directly from formula (\ref{hspec}) it is easy to see that
\begin{equation}
  \lim_{k\rightarrow 0}h_U(k)=0.
\end{equation}
Therefore using formula (\ref{zetagen}) to compute the residues and the values at the non-positive integers of $\zeta_U(s)$ for any $\Delta_U\in\mf$ is not
possible using the function (\ref{hspec}) because it does not satisfy the condition (\ref{specfzero}).
Hence we need to extract from (\ref{hspec}) the suitable function by studying the behaviour of $h_U(z)$ as $z\to 0$.

\subsection{Behaviour of $h_U(z)$ as $z\to 0$}

If we perform power series expansion in $k$ around $k=0$ of the secular equation given by (\ref{hspec}) up to first order in $k$ we obtain
\begin{equation}
    h_U(k)=2 ike^{i\alpha} \left(L \left(\cos (\alpha )-\cos (\beta )\right)-2\left(n_1 \sin (\beta )+ \sin (\alpha)\right)\right)+O\left(k^2\right).
\end{equation}
Taking into account eq. (\ref{detdu}) for any $\Delta_U\in\mf$ we can write the power series expansion above as
\begin{equation}
    h_U(k)=ik\det(D_U)+O\left(k^2\right).\label{hu1ordk}
\end{equation}
Hence for any $\Delta_U\in\mf-\mfz$ the function that satisfies the required conditions to be used in the representation of the spectral zeta function given by eq. (\ref{zetagen}) is
\begin{equation}
  \Delta_U\in\mf-\mfz\quad\Rightarrow\quad  f_U(k)=\frac{h_U(k)}{2ike^{i\alpha}}.\label{funoz}
\end{equation}
When the selfadjoint extension has a constant zero mode ($\Delta_U\in{\cal M}_F^{(0)}$) the first order in $k$ of the power expansion (\ref{hu1ordk}) is zero.
Therefore we must expand $h_U$ up to order 3 (notice from eq. (\ref{hspec}) the function $h_U$ is odd in $k$) to study the behavior at the origin:
\begin{equation}
    \det(D_U)=0\Rightarrow h_U(k)=\frac{k^3 L}{3}  \left(L\left(2 \sin (\alpha )-n_1 \sin (\beta )\right)+3\left( \cos (\alpha )+ \cos
   (\beta )\right)\right)+O(k^5).
\end{equation}
Hence, in order to obtain the function that satisfies the conditions under which (\ref{zetagen}) is valid, we must divide by an extra $k^2$ when $\Delta_U\in{\cal M}_F^{(0)}$:
\begin{equation}
  \Delta_U\in\mfz\quad\Rightarrow\quad  f^{(0)}_U(k)=\frac{h_U(k)}{2ik^3e^{i\alpha}}.\label{fusiz}
\end{equation}


\subsection{Heat kernel coefficients for $\Delta_U\in{\cal M}_F-{\cal M}^{(0)}_F$}
For this case, when $\cos (\alpha) + \cos (\beta) \neq 0$, the appropriate function is given by eq. (\ref{funoz}). Using (\ref{hspec}) we can rewrite (\ref{funoz}) for $k=ix$ as
\begin{eqnarray}
f_U(ix)&=&x e^{x L} \frac{\cos(\alpha)+\cos(\beta)}{2}\left[ 1+\frac{2}{x}\frac{\sin(\alpha)}{\cos(\alpha)+\cos(\beta)} +\frac{1}{x^2}\frac{\cos(\beta)-\cos(\alpha)}{\cos(\alpha)+\cos(\beta)}\right.\nonumber\\
&-&e^{-2 x L}\left( 1+\frac{x^{-2}(\cos(\beta)-\cos(\alpha))}{\cos(\alpha)+\cos(\beta)}-\frac{2x^{-1}\sin(\alpha)}{\cos(\alpha)+\cos(\beta)}\right)\nonumber\\
&+&\left.x^{-1}e^{-x L}\frac{4n_1\sin(\beta)}{\cos(\alpha)+\cos(\beta)}\right].\label{fuixnoz}
\end{eqnarray}
For positive $L$, the second and third line in eq. (\ref{fuixnoz}) represent exponentially damped terms as $x\to\infty$.
These terms do not contribute to the poles and to the values of $\zeta_{\Delta_U} (s)$ at non-positive integers.
Therefore, we can neglect them in the following formulas and just denote them as $e.s.t.$ Hence $\log(f_U(ix))$ will be given by
\begin{equation}
\log(f_U(ix))=\log\left(\frac{\cos(\alpha)+\cos(\beta)}{2}\right)+\log(x)+xL+\log(1+\tau_U(x)),
\end{equation}
where $\tau_U(x)$ is given by
\begin{equation}
\tau_U(x)=\frac{2}{x}\frac{\sin(\alpha)}{\cos(\alpha)+\cos(\beta)} +\frac{1}{x^2}\frac{\cos(\beta)-\cos(\alpha)}{\cos(\alpha)+\cos(\beta)} +e.s.t.
\end{equation}
Now if we take into account the series expansion
\begin{equation}
\log(1+\tau)=\sum_{n=1}^\infty(-1)^{n+1}\frac{\tau^n}{n},
\end{equation}
we can write
\begin{equation}
\log(1+\tau_U(x))=\sum_{n=1}^\infty\frac{(-1)^{n+1}}{n}\left(\frac{2}{x}\frac{\sin(\alpha)}{\cos(\alpha)+\cos(\beta)} +\frac{1}{x^2}\frac{\cos(\beta)-\cos(\alpha)}{\cos(\alpha)+\cos(\beta)}\right)^n+e.s.t.
\end{equation}
Using Newton's binomial formula we can write
\begin{equation}
\tau_U(x)^n/n=\sum_{j=0}^n\frac{\Gamma(n)2^{n-j}\sin^{n-j}(\alpha)}{\Gamma(j+1)\Gamma(n-j+1)}\frac{(\cos(\beta)-\cos(\alpha))^j}{(\cos(\alpha)+\cos(\beta))^n}x^{-(n+j)}.
\end{equation}
After reordering the double summation we obtain
\begin{eqnarray}
&&\log(1+\tau_U(x))=\sum_{m=1}^\infty b_mx^{-m},\\
&& b_m\equiv\sum_{j=0}^{[m/2]}(-1)^{m-j+1}\frac{2^{m-2j}\Gamma(m-j)\sin^{m-2j}(\alpha)}{\Gamma(j+1)\Gamma(m-2j+1)} \frac{(\cos(\beta)-\cos(\alpha))^j}{(\cos(\alpha)+\cos(\beta))^{m-j}},\label{jos1}
\end{eqnarray}
where $m=1,2,3,...$. Hence, finally we obtain the following asymptotic series for $\partial_x\log(f_U(ix))$,
\begin{equation}
\partial_x\log(f_U(ix))=L+x^{-1}-\sum_{m=1}^\infty m b_mx^{-m-1}+e.s.t.
\end{equation}
Taking into account the integral representation (\ref{zetagen}) we can write for any selfadjoint extension $\Delta_U\in{\cal M}_F-{\cal M}^{(0)}_F$,
\begin{equation}
\zeta_{\Delta_U}(s)=\frac{\sin(\pi s)}{\pi}\int_0^1 d k\cdot k^{-2s}\partial_k\log(f_U(ik))+\frac{\sin(\pi s)}{\pi}\int_1^\infty d k\cdot k^{-2s}\partial_k\log(f_U(ik)) . \label{anyselfzet}
\end{equation}
With this splitting all the information about the poles and the values of $\zeta_{\Delta_U}(s)$ at the non-positive integers is contained in the integration from $1$ to $\infty$.
Therefore, in order to perform the analytic continuation of $\zeta_{\Delta_U}(s)$ to the complex plane, we have to perform the analytic continuation of
\begin{equation}
\frac{\sin(\pi s)}{\pi}\int_1^\infty d k\cdot k^{-2s}\partial_k\log(f_U(ik)) \label{kk3.24}
\end{equation}
to the complex plane. In order to do so we must remember the following identities:
\begin{eqnarray}
\int_1^\infty dz\cdot z^{-2s}&=&\frac{1/2}{s-1/2},\\
\int_1^\infty dz\cdot z^{-2s-1}&=&\frac{1/2}{s},\\
\int_1^\infty dz\cdot z^{-2s-m-1}&=&\frac{1/2}{s+m/2}.
\end{eqnarray}
Hence, the relevant information about the analytic continuation of (\ref{kk3.24}) is contained in
\begin{equation}
\frac{\sin(\pi s)}{\pi}\int_1^\infty d k\cdot k^{-2s}\partial_k\log(f_U(ik))=\frac{\sin(\pi s)}{\pi}\left(\frac{L/2}{s-1/2}+\frac{1/2}{s} -\sum_{m=1}^{N-1} b_m\frac{m/2}{s+m/2}
+A(s)\right),
\end{equation}
where $N\in\mathbb{N}$ and $A(s)$ in the bracket of the right hand side represents a meromorphic function of $s$ analytic for $\Re s > -N/2$. The integer $N$ can be chosen as large
as we wish and details of the function $A(s)$ are irrelevant for our purposes in this section.
Using this analytic continuation, the poles of $\zeta_{\Delta_U}(s)$ can be easily computed:
\begin{eqnarray}
&&{\rm res}\left(\zeta_{\Delta_U}(s),s=1/2\right)={\rm res}\left(L\sin(\pi s)/(2\pi(s-1/2),s=1/2\right)\nonumber\\
&\Rightarrow &{\rm res}\left(\zeta_{\Delta_U}(s),s=1/2\right)=\frac{L}{2\pi},\label{reszhalf}\\
&&{\rm res}\left(\zeta_{\Delta_U}(s),s=-\frac{(2n+1)}{2}\right)={\rm res}\left(-\frac{(2n+1)b_{2n+1}\sin(\pi s)}{2\pi(s+(2n+1)/2)},s=-\frac{(2n+1)}{2}\right)\nonumber\\
&\Rightarrow &{\rm res}\left(\zeta_{\Delta_U}(s),s=-\frac{(2n+1)}{2}\right)=(-1)^{n}b_{2n+1}\frac{(2n+1)}{2\pi}, \,\,\,\,n=0,1,2,3...\label{resz-halfint}
\end{eqnarray}
Furthermore, it gives the values of $\zeta_{\Delta_U}(s)$ at the non-positive integers:
\begin{eqnarray}
&&\zeta_{\Delta_U}(0)={1\over 2}\lim_{s\rightarrow 0}\frac{\sin(\pi s)}{\pi s}=1/2,\label{zeta0}\\
&&\zeta_{\Delta_U}(-n)=-n b_{2n}\lim_{s\rightarrow -n}\frac{\sin(\pi s)}{\pi(s+n)}=(-1)^{n+1} nb_{2n},\,\,\,\,n=1,2,3...\label{zeta-n}
\end{eqnarray}
Given eqs. (\ref{reszhalf})-(\ref{zeta-n}), for any $\Delta_U\in{\cal M}_F-{\cal M}^{(0)}_F$ such that $\cos(\alpha)+\cos(\beta)\neq 0$ it is easy to compute the heat kernel coefficients using the general formulas (\ref{genhc1}) and (\ref{genhc2}). Namely, we find
\begin{eqnarray}
&&a_0=\frac{L}{2\sqrt{\pi}},\quad a_{n+1}=-\frac{4^n n! b_{2n+1}}{(2n)!\sqrt{\pi}}, \,\,\,n=0,1,2,3,...\label{hkcint-nozm}\\
&&a_{1/2}=1/2,\quad a_{n+1/2}=-\frac{b_{2n}}{(n-1)!},\,\,\,n=1,2,3,...\label{hkchalfint-nozm}
\end{eqnarray}

\subsubsection{The case of $\Delta_U\in{\cal M}_F-{\cal M}^{(0)}_F$ with $\cos(\alpha)+\cos(\beta)=0$}
For this case the appropriate function reads
\begin{eqnarray}
f^{(B)}_U(ix)&=&e^{x L} \left[ \sin(\alpha) +\frac{1}{2x}\left(\cos(\beta)-\cos(\alpha)\right)+e^{-x L}2n_1\sin(\beta)\right.\nonumber\\
&+&\left. e^{-2 x L}\left( \sin(\alpha)-\frac{1}{2x}(\cos(\beta)-\cos(\alpha))\right)\right].\label{fbuix}
\end{eqnarray}
Following the same procedure as in the general case we expand, for $\alpha \neq \pi$,
\begin{equation}
\log\left(f^{(B)}_U(ix)\right)=\log\left(\sin(\alpha)\right)+xL+\sum_{m=1}^\infty c_mx^{-m}+e.s.t.,
\end{equation}
\begin{equation}
c_m=-\frac{{\rm cotg}^m(\alpha)}{m} .
\end{equation}
Again the analytical continuation of
\begin{equation}
\frac{\sin(\pi s)}{\pi}\int_1^\infty d k\cdot k^{-2s}\partial_k\log(f_U(ik))
\end{equation}
provides the residues at the half integers and the values at the non-positive integers of $\zeta^{(B)}_{\Delta_U}(s)$,
\begin{eqnarray}
&&{\rm res}\left(\zeta^{(B)}_{\Delta_U}(s),s=1/2\right)=\frac{L}{2\pi},\\
&&{\rm res}\left(\zeta^{(B)}_{\Delta_U}(s),s=-\frac{(2n+1)}{2}\right)=(-1)^{n}c_{2n+1}\frac{(2n+1)}{2\pi},\,\,\,\,n=0,1,2,3...,\\
&&\zeta^{(B)}_{\Delta_U}(0)=0,\\
&&\zeta^{(B)}_{\Delta_U}(-n)=(-1)^{n+1}nc_{2n},\,\,\,\,n=1,2,3...
\end{eqnarray}
Once we use formulas (\ref{genhc1}) and (\ref{genhc2}) we obtain the corresponding heat kernel coefficients,
\begin{eqnarray}
&&a^{(B)}_0=\frac{L}{2\sqrt{\pi}},\quad a^{(B)}_{n+1}=-\frac{4^n n! c_{2n+1}}{(2n)!\sqrt{\pi}}, \,\,\,n=0,1,2,3,...,\\
&&a^{(B)}_{1/2}=0,\quad a^{(B)}_{n+1/2}=-\frac{c_{2n}}{(n-1)!},\,\,\,n=1,2,3,...
\end{eqnarray}
Finally, the case $\alpha = \pi$, $\beta =0$, has to be treated separately and
$$\left.\partial_x \left( \ln f_U^{(B)} (ix) \right)\right|_{\alpha = \pi} = L - \frac 1 x + e.s.t.$$
From here,
$$ {\rm res} \left( \zeta_{\Delta_U}^{(B)} (s) \right|_{\alpha =\pi}, \left. s= \frac 1 2 \right) = \frac L {2\pi} , \quad \quad
\left.\zeta_{\Delta_U}^{(B)} (0) \right|_{\alpha = \pi} = - \frac 1 2 , $$
and \begin{eqnarray}\left. a_0^{(B)} \right|_{\alpha = \pi} = \frac L { 2 \sqrt \pi} , \quad \quad \left.a_{1/2} ^{(B)} \right|_{\alpha =\pi} = - \frac 1 2,\label{heatdir}\end{eqnarray}
with all other residues and relevant values respectively heat kernel coefficients equal to zero.


\subsection{Heat kernel coefficients for $\Delta_U\in{\cal M}^{(0)}_F$}

Taking into account (\ref{mf0}) we can write (\ref{fusiz}) as
\begin{equation}
f_U^{(0)}(ix)=\frac{e^{xL}}{x}\cos(\alpha)\left[1+ \frac{\tan(\alpha)}{x}-2\frac{e^{-xL}\tan(\alpha)}{x} -e^{-2xL}\left(1-\frac{\tan(\alpha)}{x}\right)\right].
\end{equation}
Therefore
\begin{equation}
\log\left(f_U^{(0)}(ix)\right)=\log(\cos(\alpha))+xL-\log(x)+\sum_{n=1}^\infty\frac{(-1)^{n+1}\tan^n (\alpha)}{n} x^{-n}+ e.s.t.
\end{equation}
\begin{equation}
\Rightarrow\partial_x\log\left(f_U^{(0)}(ix)\right)=L-\frac{1}{x}+\sum_{n=1}^\infty(-1)^{n}\tan^n(\alpha) x^{-n-1}+ e.s.t.
\end{equation}
Hence the analytical continuation gives as before the required residues and values of $\zeta^{(0)}_{\Delta_U}(s)$,
\begin{eqnarray}
&&{\rm res}\left(\zeta^{(0)}_{\Delta_U}(s);s=1/2\right)=L/2\pi,\\
&&{\rm res}\left(\zeta^{(0)}_{\Delta_U}(s);s=-\frac{2n+1}{2}\right)=\frac{(-1)^n}{2\pi} \tan^{2n+1}(\alpha),\,\,\, n=0,1,2,3,...,\\
&&\zeta^{(0)}_{\Delta_U}(0)=-1/2,\\
&&\zeta^{(0)}_{\Delta_U}(-n)=\frac 1 2 (-1)^n\tan^{2n}(\alpha),\,\,\, n=1,2,3,...
\end{eqnarray}
To obtain the heat kernel coefficients, we must take into account that there is one zero mode in all
cases as demonstrated previously. Therefore we must add 1 to $a_{1/2}$:
\begin{eqnarray}
&& a^{(0)}_0=\frac L {2\sqrt{\pi}},\quad a^{(0)}_{1/2}=\frac 1 2 ,\label{heatneu1}\\
&& a^{(0)}_{n+1}=-\frac{4^n n! \tan^{2n+1}(\alpha)}{(2n+1)!\sqrt{\pi}},\,\,\,n=0,1,2,3,...,\label{heatneu2}\\
&& a^{(0)}_{n+1/2}=\frac 1 2 \frac{\tan^{2n}(\alpha)}{n!},\,\,\,n=1,2,3,...\label{heatneu3}
\end{eqnarray}
The heat kernel coefficients obtained above for $\Delta_U\in{\cal M}^{(0)}_F$ become singular for $\alpha=\pi/2$. In this case instead
\begin{equation}
\left.\partial_x\log\left(f_U^{(0)}(ix)\right)\right\vert_{\alpha=\pi/2}=L-\frac{2}{x}+ e.s.t.,
\end{equation}
and therefore the spectral zeta function $\left.\zeta^{(0)}_{\Delta_U}(s)\right\vert_{\alpha=\pi/2}$ will only have a residue at $s=1/2$ and non zero value at $s=0$,
\begin{eqnarray}
&&{\rm res}\left(\left.\zeta^{(0)}_{\Delta_U}(s)\right\vert_{\alpha=\pi/2};s=1/2\right)=L/2\pi,\\
&&\left.\zeta^{(0)}_{\Delta_U}(0)\right\vert_{\alpha=\pi/2}=-1.
\end{eqnarray}
Hence the only non-vanishing heat kernel coefficients are given by
\begin{eqnarray}\label{hcorners1}
&& \left. a^{(0)}_0\right\vert_{\alpha=\pi/2}=\frac L {2\sqrt{\pi}}.
\end{eqnarray}

\paragraph{The Von Neumann-Krein extension} The corresponding results for the VNK extension follow from
\begin{equation}
f_{VNK}(k)=\frac{h_{VNK}(k)}{2 i k^5 e^{i \alpha_{VNK}}}=\frac{\sin(\beta_{VNK}) }{k^{4}}(k L \sin (k L)+2 \cos (k L)-2) . \label{fvnk}
\end{equation}
Note, we divided by $k^5$ instead of the $k^3$ as in eq.~(\ref{fusiz}), this being necessary because the VNK extension has two zero modes.
The large-$k$ expansion relevant for the heat kernel coefficients reads
\begin{equation}
f_{VNK} (ik) = \frac{ 2 \left( 1-\frac{kL} 2\right)} {k^4 \sqrt{L^2+4}} e^{kL} \left( 1 + e.s.t.\right),\label{fvnkinfty}
\end{equation}
and the coefficients follow along the lines explained to be
\begin{eqnarray}
&&a^{(_{VNK})}_0=\frac L {2\sqrt{\pi}},\quad a^{(_{VNK})}_{n+1}=\frac{ n! (4/L)^{2n+1}}{(2n+1)!2\sqrt{\pi}}, \,\,\,n=0,1,2,3,...,\\
&&a^{(_{VNK})}_{1/2}=\frac 1 2,\quad a^{(_{VNK})}_{n+1/2}=\frac 1 2 \frac{(2/L)^{2n}}{n!},\,\,\,n=1,2,3,...
\end{eqnarray}
Let us stress, that in order to obtain $a_{1/2}^{(VNK)}$ we have added $+2$ to $\zeta_{VNK}^{(0)} (0)$, as is requested by having two zero modes.

\subsection{Heat kernel coefficients for common boundary conditions.}
As a check of our calculations let us compare the results found for the heat kernel coefficients with the known ones for the most common boundary conditions.
\begin{itemize}
\item {\bf Periodic boundary conditions}. The periodic boundary conditions are usually written as\footnote{When it is required that the solutions of the Laplace equation are smooth functions the periodic boundary conditions are given by the condition $\psi(0)=\psi(L)$. However square integrable solutions of the Laplace equation are not necessarily smooth. Therefore the condition $\psi(0)=\psi(L)$ does not neceessarily give rise to periodic boundary conditions. As an example it is worth to mention the case of Dirac delta potentials (see references \cite{Bordag:2011aa,Guilarte:2010xn,Munoz-Castaneda:2013yga} for recent developements in the interpretation of Dirac delta potentials as boundary conditions and infinitely thin kinks) where the condition $\psi(0)=\psi(L)$ is satisfyed but obviously the system does not satisfy periodic boundary conditions. Therefore in order to distinguish periodic boundary conditions from other types of point interactions it is necessary to include the second condition over the derivatives: $\psi'(0)=\psi'(L)$.}
\begin{equation}
\psi(0)=\psi(L);\quad \psi'(0)=\psi'(L).
\end{equation}
Equivalently we can write the following two independent equations for periodic boundary conditions
\begin{eqnarray*}
\psi(0)+i\psi'(0)&=&\psi(L)+i\psi'(L),\\
\psi(L)-i\psi'(L)&=&\psi(0)-i\psi'(0).
\end{eqnarray*}
Hence following the notation of eq. (\ref{phipm}) we can write the periodic boundary conditions in the form of (\ref{bcphipm}) as
\begin{equation}
\varphi_-(\psi)=\sigma_1\cdot\varphi_+(\psi),
\end{equation}
being $\sigma_1$ the corresponding Pauli matrix. Therefore the unitary matrix that characterizes periodic boundary conditions is given by $U_p=\sigma_1\in\mathcal{M}_F^{(0)}
\Rightarrow\,\,\alpha=\pi/2,\,\beta=\pm\pi/2,\, n_1=\mp 1$.
The heat kernel coefficients are given by (\ref{hcorners1}).
\item {\bf Dirichlet boundary condition}. The usual form of the Dirichlet boundary condition for any manifod $M$ with boundary $\partial M$ is
\begin{equation}
\left.\psi\right|_{\partial M}=0.
\end{equation}
As can be seen the normal dervatives $\left.\partial_n\psi\right|_{\partial M}$ do not enter in the boundary condition. Form eq. (\ref{dom}) the general boundary condition for those unitary operators $U\in\mathcal{M}$ such that $1\notin\sigma(U)$ can be written as
\begin{equation}
\left.\psi\right|_{\partial M}=i\frac{\mathbb{I}+U}{\mathbb{I}-U}\cdot \left.\partial_n\psi\right|_{\partial M}.\label{cayleybc1}
\end{equation}
From this last expression, it is immediate to notice that the Dirichlet boundary condition is obtained when $U=-\mathbb{I}$. Therefore the Dirichlet boundary condition is given by $U_D=-\mathbb{I}\in\mathcal{M}_F-\mathcal{M}_F^{(0)}
\Rightarrow\,\,\alpha=\pi,\,\beta=0$.
The heat kernel coefficients are given by (\ref{heatdir}).
\item {\bf Neumann boundary condition}. The usual form of the Neumann boundary condition for any manifod $M$ with boundary $\partial M$ is
\begin{equation}
\left.\partial_n\psi\right|_{\partial M}=0,
\end{equation}
where $\partial_n$  denotes the normal derivative to $\partial M$. As can be seen the boundary value $\left.\psi\right|_{\partial M}$ does not enter in the boundary condition. Form eq. (\ref{dom}) the general boundary condition for those unitary operators $U\in\mathcal{M}$ such that $-1\notin\sigma(U)$ can be written as
\begin{equation}
\left.\partial_n\psi\right|_{\partial M}=-i\frac{\mathbb{I}-U}{\mathbb{I}+U}\cdot \left.\psi\right|_{\partial M}.\label{cayleybc2}
\end{equation}
From this last expression it is immediate to notice that the Neumann boundary condition is obtained when $U=\mathbb{I}$. Therefore the Neumann boundary condition is given by
$U_N=\mathbb{I}\in\mathcal{M}_F^{(0)}\Rightarrow\,\,\alpha=\beta=0$. It is of note that in this case $\sin(\alpha)=0$. Therefore from (\ref{heatneu1})-(\ref{heatneu3}),
\begin{eqnarray}
&& a_0^{(N)}=-\frac{L}{2\sqrt{\pi}},\quad a_{1/2}^{(N)}=1/2,\\
&& a_{n+1/2}^{(N)}=0,\quad a_{n}^{(N)}=0,\,\,\, n=1,2,3,...
\end{eqnarray}
\item {\bf Robin boundary conditions}. The common expression for the family of Robin boundary conditions is given by (see for example reference \cite{Romeo:2001dd})
\begin{equation}
\left.\psi\right|_{\partial M}-g\left.\partial_n\psi\right|_{\partial M}=0,\quad g\in(-\infty,\infty).\label{robingen}
\end{equation}
For the case in which the boundary manifold $\partial M$ has several disjoint components $\partial M=\cup_i\Omega_i$ the family of Robin boundary conditions can be written as
\begin{equation}
\left.\psi\right|_{\Omega_i}-g_i\left.\partial_n\psi\right|_{\Omega_i}=0,\quad g_i\in(-\infty,\infty).\label{robingen2}
\end{equation}
The extreme values $g_i=0,\infty$ correspond to Dirichlet and Neumann boundary conditions respectively in the $i^{th}$ component of $\partial M$. Note that in the most general case the set of constants $g_i$ do not have to be the same for all the disjoint components $\Omega_i$ of $\partial M$. For $M=[0,L]$ the boundary is formed by two points and therefore it has two disjoint components. The most simple choice of Robin boundary conditions in this case is
\begin{equation}
-\psi'(0)=\tan\left(\frac{\alpha}{2}\right)\psi(0),\quad \psi'(L)=\tan\left(\frac{\alpha}{2}\right)\psi(L),\quad\alpha\in [0,\pi].\label{robin1d1}
\end{equation}
In a more compact notation we can write
\begin{equation}
\left.\tan \left( \frac{\alpha} 2 \right) \,\, \psi\right|_{\partial M}-\left.\partial_n\psi\right|_{\partial M}=0,\quad\alpha\in [0,\pi].\label{robin1d2}
\end{equation}
Taking into account eq. (\ref{cayleybc2}) and comparing it with expression (\ref{robin1d2}) the unitary operator $U_R$ for Robin boundary conditions satisfies the equation
\begin{equation}
\tan\left(\frac{\alpha}{2}\right)\mathbb{I}=-i\frac{\mathbb{I}-U_R}{\mathbb{I}+U_R}.
\end{equation}
Therefore the unitary operator that characterizes the family of Robin boundary conditions given by (\ref{robin1d1}) is given by $U_R=e^{i\alpha}\mathbb{I}$ as was firstly pointed out in references \cite{Asorey:2006pr,Asorey:2008xt}. Note that $U_R(\alpha=0)=\mathbb{I}=U_N$ and $U_R(\alpha=\pi)=-\mathbb{I}=U_D$. In the parametrization (\ref{uparam}) Robin boundary conditions correspond to $\beta=0$.  For $\alpha\in(0,\pi)$ $U_R(\alpha)\in\mathcal{M}_F-\mathcal{M}_F^{(0)}$ with $\cos(\alpha)+\cos(\beta)\neq 0$. Therefore the heat kernel coefficients for Robin boundary conditions are determined by eqs.
(\ref{hkcint-nozm}) and (\ref{hkchalfint-nozm}). From eq. (\ref{jos1}) it is easy to obtain the coefficients $b_m$ for the Robin boundary conditions:
\begin{equation}
b_m^{(R)}=\tan^m\left(\frac{\alpha}{2}\right)\sum_{j=0}^{[m/2]}(-1)^{m-j+1}\frac{2^{m-2j}\Gamma(m-j)}{\Gamma(j+1)\Gamma(m-2j+1)},\,\,\,m=1,2,3,...
\end{equation}
Using now eqs. (\ref{hkcint-nozm}) and (\ref{hkchalfint-nozm}) it is inmediate to compute the heat kernel coefficients for Robin boundary conditions to any desired order using any symbolic calculation software. As an example we show the first ten heat kernel coefficients:
\begin{eqnarray}
&&a_0^{(R)}=L/2\sqrt{\pi},\quad a_{1/2}^{(R)}=1/2,\\
&&a_1^{(R)}=-\frac{2 \tan \left(\frac{\alpha}{2}\right)}{\sqrt{\pi }},\quad a_{2}^{(R)}=-\frac{4 \tan ^3\left(\frac{\alpha}{2}\right)}{3\sqrt{\pi }},\\
&&a_3^{(R)}=-\frac{8 \tan ^5\left(\frac{\alpha}{2}\right)}{15\sqrt{\pi }},\quad a_{4}^{(R)}=-\frac{16 \tan ^7\left(\frac{\alpha }{2}\right)}{105\sqrt{\pi }},\\
&&a_{3/2}^{(R)}=\tan ^2\left(\frac{\alpha }{2}\right),\quad a_{5/2}^{(R)}=\frac{1}{2} \tan ^4\left(\frac{\alpha}{2}\right),\\
&&a_{7/2}^{(R)}=\frac{1}{6} \tan ^6\left(\frac{\alpha}{2}\right),\quad a_{9/2}^{(R)}=\frac{1}{24} \tan ^8\left(\frac{\alpha}{2}\right).
\end{eqnarray}
\end{itemize}
These results coincide with the results obtained by S. Dowker in Ref. \cite{dowk-cqg95} (equations (10) and (14) with $h_1=h_2=\tan(\alpha/2)$). More recently S. Fulling has studied the heat kernel coefficients for Robin boundary conditions in the Ref. \cite{full-2005}.

\section{The functional determinant of $\Delta_U$. Derivative at $s=0$ of the spectral zeta function}

In this section we compute the derivative of the zeta function at $s=0$ for each of the different cases considered in Sections 3.2 and 3.3.
As is well known, this derivative is a natural constituent when defining functional determinants of elliptic operators \cite{ray71-7-145}. As usual
we subtract and add back a suitable number of the asymptotic $k\to\infty$ terms of $\partial_k \log f_{\hat {\cal O}} (ik)$ in (\ref{zetagen}). In the current context
we have to subtract terms up to the order $1/k$ to make the integral well defined at $k=\infty$ once $s=0$ is set.
As a technical tool, at the start of this analysis it is convenient to consider a massive scalar field of mass $m$,
where $m$ will be sent to zero
at a suitable point of the computation. In this way we can avoid splitting the integral representing the zeta function into two pieces and the
computation becomes a little easier. The procedure is valid as in the limit $m\to 0$ the zeta function for the case with vanishing mass
is recovered. A presentation of (\ref{anyselfzet}) valid about $s=0$ is then given by
\begin{eqnarray}
\zeta_{\Delta_U} (s) &=& \frac{\sin \pi s} \pi \int\limits_m^\infty dk (k^2-m^2)^{-s} \partial_k \log \left[ \frac{2 f_U (ik) }{ke^{kL} (\cos \alpha + \cos \beta) } \right] \nonumber\\
& &+ \frac{\sin \pi s} \pi \int\limits_m^\infty dk (k^2 -m^2) ^{-s} \partial_k \log \left[ k e^{kL} \frac{ \cos \alpha + \cos \beta } 2 \right] . \nonumber\end{eqnarray}
The integral in the first line by construction is analytic about $s=0$ and its derivative at $s=0$ is trivially computed. The needed integrals in the second line are known
\cite{grad65b},
\begin{eqnarray}
\int\limits_m^\infty dk (k^2-m^2)^{-s} &=& \frac{ m^{1-2s} \Gamma (1-s) \Gamma \left( s - \frac 1 2 \right)} {2 \sqrt \pi} , \nonumber\\
\int\limits_m^\infty dk (k^2-m^2)^{-s} \,\,\frac 1 k &=& \frac{ m^{-2s} \pi} {2\sin \pi s} , \nonumber\end{eqnarray}
and
$$ \zeta_{\Delta_U} ' (0) = - \log \left| \frac{2 f_U (im)}{me^{mL} (\cos \alpha + \cos \beta)} \right| - Lm - \log m $$
is found. As $m\to 0$ we use
$$ \lim_{m\to 0} f_U (im) = L (\cos \alpha - \cos \beta ) - 2 (\sin \alpha + n_1 \sin \beta ) $$
to obtain
\begin{eqnarray}
\zeta_{\Delta_U} ' (0) = - \log \left|\frac{ 2L (\cos \alpha - \cos \beta ) - 4 (\sin \alpha + n_1 \sin \beta )}{\cos \alpha + \cos \beta } \right| . \label{det1}
\end{eqnarray}
The case treated in Section 3.2.1, for $\alpha \neq \pi$, follows along the same lines from
\begin{eqnarray}
\zeta_{\Delta_U} ^{(B)} (s) &=& \frac { \sin \pi s} \pi \int\limits_m^\infty dk (k^2-m^2)^{-s} \partial _k \log \left[ \frac{ f_U^{(B)} (ik) } { e^{kL} \sin \alpha } \right] \nonumber\\
& &+ \frac{\sin \pi s} \pi \int\limits_m^\infty dk (k^2-m^2)^{-s} \partial_k \log \left[ e^{kL} \sin \alpha \right].\nonumber\end{eqnarray}
In the limit as $m\to 0$ we obtain
$$ {\zeta_{\Delta_U} ^{(B)}}' (0) = - \log \left|\frac{ L (\cos \alpha - \cos \beta ) - 2 (\sin \alpha + n_1 \sin \beta )}{\sin \alpha } \right|.\label{detdet1}$$

For $\alpha = \pi$, $\beta =0$, instead we start with
\begin{eqnarray}
\zeta_{\Delta_U} ^{(B)} (s) \vert_{\alpha = \pi} &=& \frac{ \sin \pi s} \pi \int\limits_m^\infty dk (k^2-m^2)^{-s} \partial _k \log \left[ \frac{ f_U (ik) \vert_{\alpha = \pi} \,\, k}{e^{kL}} \right] \nonumber\\
& &+ \frac{ \sin \pi s} \pi \int\limits_m^\infty dk (k^2-m^2)^{-s} \partial _k \log \left[ \frac{e^{kL}} k \right]\nonumber\end{eqnarray}
to find
\begin{eqnarray}
{\zeta_{\Delta_U} ^{(B)}}' (0) \vert_{\alpha =\pi} = - \log (2L) .\label{detdir}\end{eqnarray}
We are left to treat the cases with a zero mode dealt with in Section 3.3. There, for $\alpha \neq \pi/2$, the starting point is
\begin{eqnarray}
\zeta_{\Delta_U} ^{(0)} (s) &=& \frac{\sin \pi s} \pi \int\limits_m^\infty dk (k^2-m^2)^{-s} \partial _k \log \left[ \frac{ f_U^{(0)} (ik) \,\, k} {e^{kL} \cos \alpha } \right] \nonumber\\
& &+ \frac{\sin \pi s } \pi \int\limits_m^\infty dk (k^2-m^2)^{-s} \partial _k \log \left[ \frac{ e^{kL} \cos \alpha }  k\right] \nonumber\end{eqnarray}
leading to
$$ {\zeta_{\Delta_U} ^{(0)}}' (0) = - \log \left| \frac{L (2 \cos \alpha + L \sin \alpha )} {\cos \alpha } \right|.\label{detneu}$$

For $\alpha = \pi /2$ instead
\begin{eqnarray}
\zeta_{\Delta_U} ^{(0)} (s) \vert_{\alpha = \pi /2} &=& \frac{\sin \pi s} \pi \int\limits_m^\infty dk (k^2-m^2)^{-s} \partial _k \log \left[\frac{ f_U^{(0)} (ik)\vert_{\alpha = \pi/2} k^2} {e^{kL} } \right] \nonumber\\
& & + \frac{ \sin \pi s} \pi \int\limits_m^\infty dk (k^2-m^2)^{-s} \partial _k \log \left[ \frac{e^{kL}} {k^2} \right] ,\nonumber\end{eqnarray}
leading to
\begin{eqnarray}
{\zeta_{\Delta_U} ^{(0)} } ' (0) = - 2 \log L. \label{detper}
\end{eqnarray}

The expression obtained for ${\zeta_{\Delta_U} ^{(0)}}' (0)$ does not allow one to compute the derivative of the spectral zeta function at $s=0$ for the VNK extension by just replacing $\{\alpha,\beta\}\mapsto\{\alpha_{VNK},\beta_{VNK}\}$ as it does produce an undefined answer. The reason is that there are two zero modes and the formulas
have to be adapted; see eqs.~(\ref{fvnk}) and (\ref{fvnkinfty}). From the large-$k$ expansion (\ref{fvnkinfty}) and from
 $ f_{VNK}(i m)\simeq-L^4/(6\sqrt{L^2+4})+O\left(m\right)$ the computation explained above leads to the following expression for ${\zeta_{VNK} ^{(0)}}' (0) $:
\begin{equation}
{\zeta_{VNK} ^{(0)}}' (0) =-\log\left\vert \frac{\left.f_{VNK}(im)\right\vert_{m\rightarrow 0}}{-L/\sqrt{L^2+4}} \right\vert=-\log\left( \frac{L^3}{6}\right).
\end{equation}

These results can be confronted with the easily computed answers for periodic, Dirichlet and Neumann boundary conditions.

For Dirichlet boundary conditions the spectrum is $\lambda_n = (\pi n/L)^2$, $n\in\mathbb{N}$, with associated zeta function
$\zeta_{Dir} (s) = (\pi /L)^{-2s} \zeta_R (2s)$. This gives $\zeta_{Dir} ' (0) = - \log (2L)$ in agreement with (\ref{detdir}).

For Neumann boundary conditions the spectrum is as above but with zero included. For the determinant the answers therefore again reads $\zeta_{Neu} ' (0) = - \log (2L)$, which agrees with (\ref{detneu}), once
$\alpha =\beta =0$ has been put.

Finally, for periodic boundary conditions the spectrum is $\lambda_n = (2\pi n/L)^2$, $n\in\mathbb{Z}$, with associated zeta function (zero mode excluded)
$\zeta_{per} (s) = 2 (2\pi/L)^{-2s} \zeta _R (2s)$. This shows $\zeta_{per} ' (0) = - 2 \log L$, again in agreement with (\ref{detper}).

As a specific new result, Robin boundary conditions as described above follow from (\ref{det1}) as
$$\zeta _{\Delta_{U_R}} ' (0) = - \log \left( 2 \tan \left( \frac \alpha 2 \right) \left( L \tan \left( \frac \alpha 2 \right) +2 \right) \right).$$

\section{Conclusions}
In this article we have analyzed the spectral zeta function resulting from the Laplacian on the interval $[0,L]$ for the case when strongly consistent
selfadjoint extensions and the Von Neumann-Krein extension are applied. Contour integral representations for the zeta functions are obtained for this class of selfadjoint extensions. These are used
to compute leading heat kernel coefficients and the functional determinant in this context. Our results agree with known results for standard boundary conditions
like Dirichlet, Neumann and periodic. The generalisation of these results to a scalar quantum field theory in
$D+1$ spacetime confined between two $D-1$ dimensional plane parallel plates is straightforward for the heat kernel coefficients
due to the factorization properties of the heat kernel in the same way as it is done in ref. \cite{asor13-874-852}.

The current article represents the start of further investigations into the details of heat kernel coefficients. Heat kernel coefficients are usually represented
in terms of geometric invariants with universal multipliers depending on the boundary condition. The question arises how the multipliers depend on the chosen selfadjoint
extension. In order to get some nontrivial boundary geometry involved a similar computation should be done for balls along the lines of
\cite{bord96-37-895,bord96-179-215,bord96-182-371}, where choosing general selfadjoint extensions will lead to different combinations of Bessel functions. Furthermore,
following \cite{jeff12-45-345201}, surfaces of revolution are possible candidates to analyze how different selfadjoint extensions impact spectral functions.

Finally, the presented analysis could also be done for selfadjoint extensions that allow for finitely many negative eigenvalues by using a variation of the current procedure \cite{kirs04-37-4649}.
(Note, that the results presented in this article actually remain valid beyond strongly consistent selfadjoint extensions as long as the
eigenvalues are positive!)
We believe that this kind of selfadjoint extensions could provide a natural mechanism for inflation in cosmological models with compact extra dimensions with boundary (Kaluza Klein cosmology or RS-type scenarios), where the dark energy is interpreted as the quantum vacuum of a fundamental scalar. In this scenario the inflationary phase is produced by the existence of negative energy modes the existence of which is strongly dependent on the size of the compact extra dimension with boundary.

\section*{Acknowledgement}
We acknowledge the support from DFG in project  BO1112-18/1. We also acknowledge M. Asorey for valuable discussions, and S. Dowker and the referee of Lett. Math. Phys. for fruitful comments.

\appendix
\section{Some remarks on the AIM formalism}
For completeness  we would like to introduce some basic notions about the AIM formalism. For simplicity we will restrict ourselves to the case of the selfadjoint extensions of the Laplace operator over the interval $[0,L]$. The AIM formalism establishes a one-to-one correspondence between the abstract set $\mathcal{M}(L)$ of selfadjoint extensions of the Laplace operator over the finite line\footnote{In this appendix we will make a small change of notation in that we specify the $L$ dependence of the space of selfadjoint extensions of the Laplace operator over the finite interval $[0,L]$. Of course for a pair of lengths $L_1,L_2>0$, $\mathcal{M}(L_1)\simeq\mathcal{M}(L_2)\equiv\mathcal{M}$ but it is important for the moment to keep this dependence in mind.} $[0,L]$
and the unitary group $U(2)$
\begin{eqnarray*}
s_L&:&U(2)\longrightarrow\mathcal{M}(L).
\end{eqnarray*}
The isomorphism is unique for each value of $L>0$. The physical realisation of an abstract selfadjoint extension in $\mathcal{M}(L)$ is a boundary condition that determines the dynamics of the free particle on the interval $[0,L]$ together with the equation of motion given by the Laplace operator. Therefore an abstract selfadjoint extension acquires a physical meaning as a unitary operator $s_L^{-1}(\Delta_U)$ defining a boundary condition. Hence from a physical point of view what is physically meaningful is not the abstract set $\mathcal{M}(L)$ but its inverse image through $s_L$, i. e. $s_L^{-1}\left(\mathcal{M}(L)\right)\simeq U(2)$. With this image in mind, and knowing that the length $L$ of the interval plays a crucial role in physical problems such as the Casimir effect (see for example \cite{asor13-874-852}), one must assume that the basic object is the group $U(2)$. In this picture the space $\mathcal{M}_F(L)$ is defined as
\begin{equation}
\mathcal{M}_F(L)\equiv \{U\in U(2)\,\mid\, q_{L^\prime}(U)\in\mathcal{M}_{NN}(L^\prime)\,\,\forall\,\, L^\prime>0\},
\end{equation}
where $\mathcal{M}_{NN}(L)\subset\mathcal{M}(L)$ is the subset of non-negative selfadjoint extensions for a given length of the interval $L$. In this sense the set $\mathcal{M}_F$ as a set of boundary conditions is independent of the length of the interval, and it is stable under $s_L^{-1}$ for any $L>0$. Keeping boundary conditions defined by $U(2)$ matrices as the basic objects, statements concerning all values of $L$ or particular values of $L$ made throughout this paper can be reformulated in terms of the isomorphisms $s_L$. In particular, concerning the VNK extension, one should say that ``$s_L^{-1}(\Delta_{VNK})\notin\mathcal{M}_F$ for any $L>0$''.

A more geometric picture can be obtained by constructing a bundle structure over the set of selfadjoint extensions of the Laplace operator over the finite interval. The base space is given by the positive real numbers $\mathbb{R}^+$, the total space will be given by $E=\{\mathcal{M}(L)\}_{L\in\mathbb{R}^+}$,
\begin{equation*}
\pi: E\longrightarrow \mathbb{R}^+,
\end{equation*}
and the fiber is given by $F=\pi^{-1}(L)=\mathcal{M}(L)\simeq U(2)$. In this picture the abstract selfadjoint extensions $\mathcal{M}$ are nothing but sections of the fiber bundle just introduced. Obviously this bundle has a natural structure of principal bundle with structural group $U(2)$ acting on $E$ naturally through the group isomorphisms $s_L$. The VNK section would be the only section $w_{VNK}:\mathbb{R}^+\longrightarrow E$ that for every value of $L\in \mathbb{R}^+$ gives rise to a selfadjoint extension with two zero modes, and the space $\mathcal{M}_F$ should be defined as those constant sections (the sections such that $w(L)=(L,U)$ with $U\in U(2)$ independent of $L$) whose image is in $\lbrace\mathcal{M}_{NN}(L)\rbrace_{L\in\mathbb{R}^+}$.


\begin{thebibliography}{10}

\bibitem{asor13-874-852}
M.~Asorey and J.M. Munoz-Castaneda.
\newblock {Attractive and Repulsive Casimir Vacuum Energy with General Boundary
  Conditions}.
\newblock {\em Nucl.Phys.}, B874:852--876, 2013.

\bibitem{aso-jpa07}
M.~Asorey, D.~Garcia-Alvarez, and J.M. Munoz-Castaneda.
\newblock {Vacuum Energy and Renormalization on the Edge}.
\newblock {\em J.Phys.}, A40:6767--6776, 2007.

\bibitem{zett05b}
A.~Zettl.
\newblock {\em Sturm-Liouville Theory, Mathematical Surveys and Monographs}.
\newblock American Mathematical Society, 2005.

\bibitem{muca-phd09}
J.M. Munoz-Castaneda.
\newblock {Boundary effects in quantum field theory (PhD dissertation, in
  spanish)}.
\newblock {\em Zaragoza Univ.}, October 2009.

\bibitem{perez-pardo}
J.~M. {P{\'e}rez-Pardo}.
\newblock {On the Theory of Self-Adjoint Extensions of the Laplace-Beltrami
  Operator, Quadratic Forms and Symmetry}.
\newblock {\em ArXiv e-prints}, August 2013.

\bibitem{asor05-20-1001}
M.~Asorey, A.~Ibort, and G.~Marmo.
\newblock {Global theory of quantum boundary conditions and topology change}.
\newblock {\em Int.J.Mod.Phys.}, A20:1001--1026, 2005.

\bibitem{eliz94b}
{E. Elizalde, S.D. Odintsov, A.~Romeo, A.A. Bytsenko, and S.~Zerbini}.
\newblock {\em Zeta Regularization Techniques with Applications}.
\newblock World Scientific, Singapore, 1994.

\bibitem{gilk95b}
P.B. Gilkey.
\newblock {\em Invariance Theory, the Heat Equation and the Atiyah-Singer Index
  Theorem}.
\newblock CRC Press, Boca Raton, 1995.

\bibitem{gilk04b}
P.B. Gilkey.
\newblock {\em Asymptotic formulae in spectral geometry}.
\newblock Chapman \& Hall/CRC, Boca Raton, 2004.

\bibitem{kirs02b}
K.~Kirsten.
\newblock {\em Spectral Functions in Mathematics and Physics}.
\newblock Chapman\&Hall/CRC, Boca Raton, FL, 2002.

\bibitem{vass03-388-279}
D.V. Vassilevich.
\newblock {Heat kernel expansion: User's manual}.
\newblock {\em Phys.Rept.}, 388:279--360, 2003.

\bibitem{ray71-7-145}
D.B. Ray and I.M. Singer.
\newblock {R-torsion and the Laplacian on Riemannian manifolds}.
\newblock {\em Advances in Math.}, 7:145--210, 1971.

\bibitem{blau88-310-163}
S.~Blau, M.~Visser, and A.~Wipf.
\newblock {Zeta Functions and the Casimir Energy}.
\newblock {\em Nucl.Phys.}, B310:163, 1988.

\bibitem{bord09b}
M.~Bordag, G.L. Klimchitskaya, U.~Mohideen, and V.M. Mostepanenko.
\newblock {\em {Advances in the Casimir effect}}.
\newblock Oxford University press, 2009.

\bibitem{bord01-353-1}
M.~Bordag, U.~Mohideen, and V.M. Mostepanenko.
\newblock {New developments in the Casimir effect}.
\newblock {\em Phys.Rept.}, 353:1--205, 2001.

\bibitem{byts96-266-1}
A.~A. Bytsenko, G.~Cognola, L.~Vanzo, and S.~Zerbini.
\newblock {Quantum fields and extended objects in space-times with constant
  curvature spatial section}.
\newblock {\em Phys.Rept.}, 266:1--126, 1996.

\bibitem{dowk76-13-3224}
J.S. Dowker and R.~Critchley.
\newblock {Effective Lagrangian and Energy Momentum Tensor in de Sitter Space}.
\newblock {\em Phys.Rev.}, D13:3224, 1976.

\bibitem{dowk78-11-895}
J.S. Dowker and G.~Kennedy.
\newblock {Finite Temperature and Boundary Effects in Static Space-Times}.
\newblock {\em J.Phys.}, A11:895, 1978.

\bibitem{hawk77-55-133}
S.W. Hawking.
\newblock {Zeta Function Regularization of Path Integrals in Curved
  Space-Time}.
\newblock {\em Commun.Math.Phys.}, 55:133, 1977.

\bibitem{milt01b}
K.A. Milton.
\newblock {\em The Casimir Effect: Physical Manifestations of Zero-Point}.
\newblock River Edge, USA: World Scientific, 2001.

\bibitem{alsi-jot80}
A.~Alonso and B.~Simon.
\newblock The {B}irman-{K}re\u\i n-{V}ishik theory of selfadjoint extensions of
  semibounded operators.
\newblock {\em J. Operator Theory}, 4(2):251--270, 1980.

\bibitem{teschl-am10}
M.~S. Ashbaugh, F.~Gesztesy, M.~Mitrea, and G.~Teschl.
\newblock Spectral theory for perturbed krein laplacians in nonsmooth domains.
\newblock {\em Advances in Mathematics}, 223(4):1372 -- 1467, 2010.

\bibitem{teschl-12}
M.~S. {Ashbaugh}, F.~{Gesztesy}, M.~{Mitrea}, R.~{Shterenberg}, and
  G.~{Teschl}.
\newblock {A Survey on the Krein-von Neumann Extension, the corresponding
  Abstract Buckling Problem, and Weyl-Type Spectral Asymptotics for Perturbed
  Krein Laplacians in Nonsmooth Domains}.
\newblock {\em ArXiv e-prints}, March 2012.

\bibitem{kirs03-308-502}
K.~Kirsten and A.~J. McKane.
\newblock {Functional determinants by contour integration methods}.
\newblock {\em Annals Phys.}, 308:502--527, 2003.

\bibitem{seel68-10-288}
R.T. Seeley.
\newblock {Complex powers of an elliptic operator, Singular Integrals, Chicago
  1966}.
\newblock {\em Proc. Sympos. Pure Math.}, 10:288--307, 1968.

\bibitem{Bordag:2011aa}
M.~Bordag and J.M. Munoz-Castaneda.
\newblock {Quantum vacuum interaction between two sine-Gordon kinks}.
\newblock {\em J.Phys.}, A45:374012, 2012.

\bibitem{Guilarte:2010xn}
J.~Mateos~Guilarte and J.~M. Munoz-Castaneda.
\newblock {Double-delta potentials: one dimensional scattering. The Casimir
  effect and kink fluctuations}.
\newblock {\em Int.J.Theor.Phys.}, 50:2227--2241, 2011.

\bibitem{Munoz-Castaneda:2013yga}
J.~M. Munoz-Castaneda, J.~Mateos~Guilarte, and A.~Moreno~Mosquera.
\newblock {Quantum vacuum energies and Casimir forces between partially
  transparent $\delta$-function plates}.
\newblock {\em Phys.Rev.}, D87(10):105020, 2013.

\bibitem{Romeo:2001dd}
A.~Romeo and A.~A. Saharian.
\newblock {Vacuum densities and zero point energy for fields obeying Robin
  conditions on cylindrical surfaces}.
\newblock {\em Phys.Rev.}, D63:105019, 2001.

\bibitem{Asorey:2006pr}
M.~Asorey, D.~Garcia-Alvarez, and J.M. Munoz-Castaneda.
\newblock {Casimir Effect and Global Theory of Boundary Conditions}.
\newblock {\em J.Phys.}, A39:6127--6136, 2006.

\bibitem{Asorey:2008xt}
M.~Asorey and J.M. Munoz-Castaneda.
\newblock {Vacuum Boundary Effects}.
\newblock {\em J.Phys.}, A41:304004, 2008.

\bibitem{dowk-cqg95}
J.S. Dowker.
\newblock {Robin conditions on the Euclidean ball}.
\newblock {\em Class.Quant.Grav.}, 13:585--610, 1996.

\bibitem{full-2005}
Stephen~A. Fulling.
\newblock {Local spectral density and vacuum energy near a quantum graph
  vertex}.
\newblock {\em arXiv:math/0508335}, 2005.

\bibitem{grad65b}
I.S. Gradshteyn and I.M. Ryzhik.
\newblock {\em Table of integrals, series, and products}.
\newblock Academic Press, New York, 1965.

\bibitem{bord96-37-895}
M.~Bordag, E.~Elizalde, and K.~Kirsten.
\newblock {Heat kernel coefficients of the Laplace operator on the
  D-dimensional ball}.
\newblock {\em J.Math.Phys.}, 37:895--916, 1996.

\bibitem{bord96-179-215}
M.~Bordag, B.~Geyer, K.~Kirsten, and E.~Elizalde.
\newblock {Zeta function determinant of the Laplace operator on the
  D-dimensional ball}.
\newblock {\em Commun.Math.Phys.}, 179:215--234, 1996.

\bibitem{bord96-182-371}
M.~Bordag, Klaus Kirsten, and J.S. Dowker.
\newblock {Heat kernels and functional determinants on the generalized cone}.
\newblock {\em Commun.Math.Phys.}, 182:371--394, 1996.

\bibitem{jeff12-45-345201}
T.~D. Jeffres, K.~Kirsten, and T.~Lu.
\newblock {Zeta Function on Surfaces of Revolution}.
\newblock {\em J. Phys.A}, 45(34):345201, 2012.

\bibitem{kirs04-37-4649}
K.~Kirsten and A.~J. McKane.
\newblock {Functional determinants for general Sturm-Liouville problems}.
\newblock {\em J.Phys.}, A37:4649--4670, 2004.

\end{thebibliography}

\end{document}